\documentclass{amsart}
\usepackage{graphicx}    
\usepackage{url}
\usepackage{latexsym,amssymb}
\usepackage[textsize=tiny]{todonotes}

\usepackage[margin=1in]{geometry}

\usepackage{setspace}
\usepackage{amsmath}

\usepackage{latexsym}
\usepackage{enumitem}
\usepackage{amsmath}
\usepackage{accents}

\usepackage{dsfont}
\usepackage{bbm}
\usepackage{bm}
\usepackage{epsfig}
\usepackage{subfigure}
\usepackage{color}
\usepackage{natbib}

\newtheorem{thm}{Theorem}[section]

\newtheorem{example}[thm]{Example}
\newtheorem{remark}[thm]{Remark}

\newtheorem{proposition}[thm]{Proposition}
\numberwithin{equation}{section}

\renewenvironment{proof}{\noindent {\bf Proof.\ }}{\hfill{\rule{2mm}{2mm}}}

\newcommand{\dd}{\mathrm{d}}
\newcommand{\ind}{\mathds{1}} 
\newcommand{\cd}{\stackrel{d}{\rightsquigarrow}}

\newcommand{\cboxed}[2]{\colorbox{#1!20}{\ensuremath{\displaystyle #2}}}

\author{Jiaying Gu}
\address{Department of Economics, University of Toronto}
\email{jiaying.gu@utoronto.ca}
\author{Nikolaos Ignatiadis}
\address{Department of Statistics and Data Science Institute, University of Chicago}
\email{ignat@uchicago.edu}
\author{Azeem M. Shaikh}
\address{Department of Economics, University of Chicago}
\email{amshaikh@uchicago.edu}

\begin{document}
\bibliographystyle{econometrica}
\title[Reasonable uncertainty]{Reasonable uncertainty: Confidence intervals in empirical Bayes discrimination detection}
\onehalfspacing

\begin{abstract}
We revisit empirical Bayes discrimination detection, focusing on uncertainty arising from both partial identification and sampling variability. While prior work has mostly focused on partial identification, we find that some empirical findings are not robust to sampling uncertainty. To better connect statistical evidence to the magnitude of real-world discriminatory behavior, we propose a counterfactual odds-ratio estimand with a attractive properties and interpretation. Our analysis reveals the importance of careful attention to uncertainty quantification and downstream goals in empirical Bayes analyses.
\end{abstract}

\maketitle

\section{Introduction}

Empirical Bayes ~\citep{robbins1956empirical, efron2019bayes} methods are increasingly popular in applied research.  Prominent examples covering a diverse array of applications include studies by \cite{rozema2019good} on police behavior, \cite{wernerfelt2022estimating} on advertising treatment effects, \cite{gu2022ranking} on journal ratings, \cite{NBERw31192} on managerial productivity effects, and \cite{coey2022empirical} on online controlled experiments.  The growing importance of empirical Bayes methods is further highlighted by \cite{walters2024empirical}, who surveys applications in labor economics.   A distinguishing feature of these and other empirical Bayes analyses is that they rarely include uncertainty quantification for posterior estimands.  

In this paper, we revisit this methodological shortcoming in the context of a compelling application of empirical Bayes described in \citet{kline2021reasonable} to discrimination detection based on correspondence experiments.  Their analysis emphasizes the role of uncertainty stemming from partial identification, but, with a few notable exceptions that we describe further below in Section \ref{sec:optimi}, \citeauthor{kline2021reasonable} primarily report point estimates for their posterior estimands without further incorporating sampling uncertainty.  Our discussion, by contrast, highlights the way in which partial identification and sampling uncertainty for posterior estimands are intertwined and naturally addressed in concert.  In the course of doing so, we draw attention to some new results on confidence intervals for empirical Bayes analyses and introduce some novel methods exploiting inference methods recently developed for other problems.  

We apply these methods to data from one of the three correspondence experiments analyzed by \citeauthor{kline2021reasonable}, specifically the study by \cite{arceo-gomez2014race} of race and gender discrimination in Mexico City. After doing so, we find that some of the empirical conclusions in \citet{kline2021reasonable} concerning this data prove substantially more robust than others.  In this way, our analysis demonstrates the importance of accounting for sampling uncertainty in empirical Bayes discrimination analysis.  We further show that this remains true for an alternative estimand based on an odds ratio that we argue may be preferred relative to the one considered in \citet{kline2021reasonable} for several different reasons.  Through these contributions, we hope to make it routine to report confidence intervals alongside point estimates for empirical Bayes estimands and, as called for by \citet{imbens2022comment}, to encourage further research on statistical inference accompanying empirical Bayes analyses.

The remainder of our paper is organized as follows.  In Section \ref{sec:setup}, we first review the key ingredients of an empirical Bayes analysis and then specialize our discussion to the setting in \citet{kline2021reasonable}.  In Section \ref{sec:optimi}, we discuss partial identification and sampling uncertainty through the lens of four different optimization problems.  This discussion motivates a particular approach to account for uncertainty described in Section \ref{sec:$F$-Localization} based on \citet{ignatiadis2022confidence}.  Other methods to account for uncertainty are described in Section \ref{sec:further_methods}, including a novel application of \citet{fang2023inference}.  Along the way, we apply each of these methods to re-assess some empirical conclusions in \cite{kline2021reasonable}.  Finally, in Section \ref{sec:estimand}, we introduce and discuss our alternative estimand, and apply these methods to it as well.

\section{Setup and Notation} \label{sec:setup}

A canonical empirical Bayes analysis~\citep{robbins1956empirical, efron2019bayes, ignatiadis2022confidence} consists of three ingredients: 
\begin{itemize}[leftmargin=*]
    \item Data from multiple related units $Z_1,\dotsc,Z_n \in \mathcal{Z}$ drawn independently based on a known likehood, $Z_i \sim p(\cdot \mid \theta_i)$, where $\theta_i$ is the parameter of interest for the $i$-th unit.
    \item A structural distribution $G$ describing the ensemble of parameters $\theta_i$ via $\theta_i \sim G$ and $G \in \mathcal{G}$, where $\mathcal{G}$ is a class of distributions.
    \item An estimand $\theta(G;z)$ that permits an oracle decision maker with knowledge of the ensemble $G$ to make an optimal decision regarding a unit with observed data $z$. 
\end{itemize}
The empirical Bayesian has no knowledge of the ensemble $G$, yet seeks to mimic the oracle decision maker by learning from indirect evidence, i.e., by using the observed $Z_1,\dotsc,Z_n$ to learn about $G$. The connection between the observed data and the unknown ensemble is established through the marginal density of the $Z_i$,
\begin{equation}
\label{eq:marginal_density}
f_G(z) = \int p(z \mid \theta) \dd G(\theta).
\end{equation}
In the setting considered by \citet{kline2021reasonable},

\begin{itemize}[leftmargin=*]
    \item Each unit $i=1,\dotsc,n$ is a job. The experimenter sends out $L$ fictitious job applications from each of two groups, labeled $a$ and $b$, and records the number of callbacks for each group, $Z_i = (C_{ai}, C_{bi}) \in \mathcal{Z}=\{0,\dotsc,L\}^2$. The likelihood is modeled as a bivariate binomial, i.e., for $z=(c_a,c_b)$,
    \begin{equation}
    \label{eq:bivariate_binomial}
     p(z \mid \theta) =  \binom{L}{c_a} \binom{L}{c_b}\,  p_a^{c_a}(1-p_a)^{L-c_a} p_b^{c_b} (1-p_b)^{L-c_b},
    \end{equation}
    where $\theta = (p_a,p_b)$ are the callback probabilities for groups $a$ and $b$, respectively. For such discrete data, the marginal density $f_G$ in~\eqref{eq:marginal_density} is a probability mass function (i.e., a density with respect to the counting measure).
    \item The structural distribution $G$ is a distribution over the unit cube $[0,1]^2$ and $\mathcal{G}$ is the class of all distributions over $[0,1]^2$, i.e., no further restrictions are imposed on $G$. 
    \item The estimand of interest is the posterior probability that a job with callback pattern $z$ discriminates against group $b$ (i.e., favors group $a$ over $b$),
    \begin{equation}
    \theta^{\text{discr}}(G;z) := \mathbb P_{G}[p_a > p_b \mid Z=z].
    \label{eq:discrimination_estimand}
    \end{equation}
\end{itemize}
\citeauthor{kline2021reasonable} demonstrate using data from three different correspondence experiments how empirical Bayes provides a principled way to detect discrimination patterns across jobs in this type of setting. There are, however, two important sources of uncertainty in such an analysis. 
First, in some empirical Bayes problems, such as the bivariate binomial model in~\eqref{eq:bivariate_binomial} described above, even if we had precise knowledge of $f_G$, we could not recover $G$ uniquely. In other words, $G$ is only partially identified. Second, in practice, we do not know $f_G$ either and must estimate it from data.  

Before proceeding, we note that we confine our analysis below to data from one of the three correspondence experiments analyzed by \citeauthor{kline2021reasonable}, specifically the study by \citet{arceo-gomez2014race} of gender discrimination, but the same considerations apply equally well to the other two correspondence experiments they analyze, namely \citet{bertrand2004are} and \citet{nunley2015racial}.

\section{On partial identification and shape-constrained GMM} \label{sec:optimi}

In this section, we discuss four different optimization problems that will help us not only explain the way in which \citeauthor{kline2021reasonable} address the partial identification issue, but also how sampling variability and partial identification are intertwined. 
The optimization problems each seek to minimize the estimand $\theta(\widetilde{G};z)$ over all distributions $\widetilde{G}$, but are distinguished by different additional constraints.\footnote{The reader should keep the estimand~\eqref{eq:discrimination_estimand} in mind, but the discussion applies to any estimand $\theta(G;z)$.}  To solve each optimization problem, we discretize \smash{$\widetilde{G}$} on a two-dimensional grid with $K^2$ points; we defer a more detailed discussion of computational issues to Section~\ref{subsec:optimization_issues}.
\vspace{.1cm}
\begin{equation}     
    \begin{aligned}     
    & \underset{\widetilde{G} \in \mathcal{G}}{\text{minimize}}
    & & \theta(\widetilde{G};z) \quad \text{subject to one of:} \\     
    & & & (i) \:\: \cboxed{green}{f_{\widetilde{G}} = f_G}, \quad
    (ii) \:\: \cboxed{gray}{f_{\widetilde{G}} = \bar{f}}, \quad         
    (iii) \:\: \cboxed{blue}{f_{\widetilde{G}} = \bar{f}^{\mathrm{proj}}}, \quad         
    (iv) \:\: \cboxed{red}{J_n(f_{\widetilde{G}}, \bar{f}) \leq \kappa}.
    \end{aligned}     
    \label{eq:opt} 
\end{equation}
Below, we will explain each of these optimization problems in turn and their constraints, defining all required quantities along the way.

Optimization problem (i) represents an idealized benchmark: if we knew the true marginal density $f_G$, what would be the smallest possible value of $\theta(\widetilde{G};z)$ among all $\widetilde{G}$ that are consistent with this density? Though one could also maximize, we focus on the minimum as, for the discrimination probability $\theta^{\text{discr}}(G;z)$, it represents the most conservative value that is compatible with the true marginal density. Optimization problem (i) captures the fundamental partial identification challenge. We next turn to problems (ii)--(iv) that also capture issues that stem from the fact that $f_G$ is not known and must be estimated.

In optimization problems (ii) and (iii) the true density $f_G$ is replaced by an estimate. Problem (ii) uses the empirical frequencies $\bar{f}(z) = \sum_{i=1}^n \ind(Z_i = z)/n$, which provide a natural estimate for discrete data.\footnote{For continuous data, estimation of $f_G$ would require more sophisticated density estimation techniques.} \citeauthor{kline2021reasonable} pursue precisely this approach to compute estimates of lower bounds in their application to the \citet{bertrand2004are} experiment. Problem (ii) may, however, be infeasible: there may be no distribution \smash{$\widetilde{G}$} whose implied marginal density exactly matches $\bar{f}$. This situation can occur due to sampling variability in $\bar{f}$, or due to misspecification of the bivariate binomial model. Indeed, infeasibility occurs in two of the three empirical examples in \citet{kline2021reasonable} and is a well understood phenomenon in the related univariate binomial problem~\citep{wood1999binomial}.

To address infeasibility, \citeauthor{kline2021reasonable} introduce the following generalized method of moments (GMM) problem:
\vspace{.1cm}
\begin{equation}
    \begin{aligned}
    & \underset{\widetilde{G} \in \mathcal{G}}{\text{minimize}}
    & & J_n(f_{\widetilde{G}}, \bar{f}),\;\;\;\;\;\;J_n(f, \bar{f}) := n  (f - \bar{f})^\intercal \widehat{W}(f - \bar{f}),
    \end{aligned}
    \label{eq:J_functional}
\end{equation}
where \smash{$\widehat{W}$} is a weighting matrix computed in a first-stage GMM step. Let \smash{$G^{\mathrm{proj}}$} be the solution to~\eqref{eq:J_functional}, $\bar{f}^{\mathrm{proj}} = f_{G^{\mathrm{proj}}}$ the implied marginal density, and $J_n^{\mathrm{opt}} = J_n(\bar{f}^{\mathrm{proj}}, \bar{f})$ the optimal value of~\eqref{eq:J_functional}.  Optimization problem (iii) then replaces $\bar{f}$ in the constraint for problem (ii) with $\bar{f}^{\text{proj}}$. Unlike problem (ii), problem (iii) is always feasible by construction, but still ignores sampling variability.

\begin{figure}
    \centering
    \includegraphics[width=0.5\textwidth]{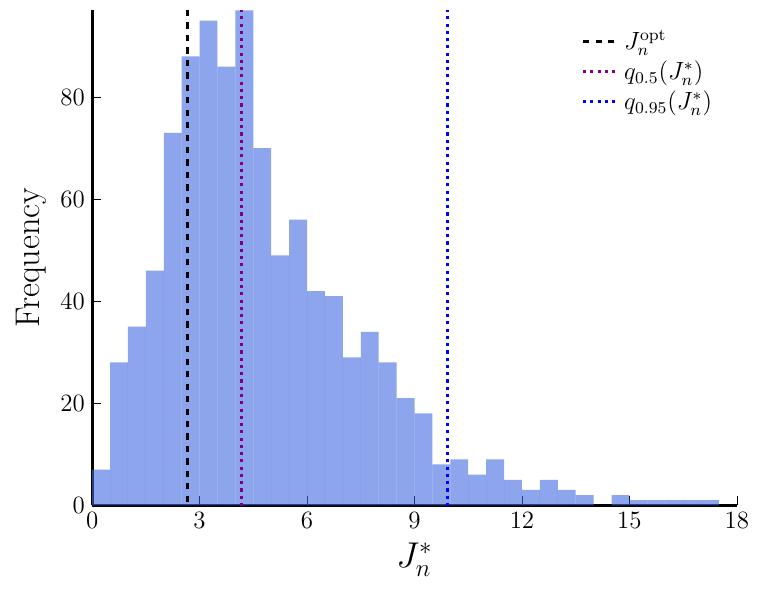}
    \caption{CNS bootstrap distribution of $J_n^{\mathrm{opt}}$ for the AGCV dataset. Based on $1,000$ replicates. }
    \label{fig:bootstrap}
\end{figure}

Following common practice for empirical Bayes analyses (as discussed in the introduction), 
\citeauthor{kline2021reasonable} only report point estimates of their lower bounds for the posterior discrimination estimand in~\eqref{eq:discrimination_estimand} without further incorporating sampling uncertainty, that is, they report the objective value of optimization problem~\ref{eq:opt}(iii) (or (ii) when feasible).  We emphasize, however, that \citeauthor{kline2021reasonable} are aware of sampling uncertainty more generally and address it in some other parts of their analysis.  They propose, for example, a shape-constrained bootstrap scheme, following~\citet{chernozhukov2023constrained} (CNS) for goodness-of-fit testing based on the distribution of $J_n^{\text{opt}}$ (the minimum value of the GMM statistic). Beyond goodness-of-fit testing, \citeauthor{kline2021reasonable} also adapt the bootstrap scheme of CNS to test null hypotheses such as $\mathbb P_G[p_a \neq p_b] = 0$ or $\mathbb P_G[p_a > p_b]=0$.

Figure~\ref{fig:bootstrap} shows the bootstrap distribution\footnote{
We generate the bootstrap samples by directly rerunning the reproduction code of~\citet{kline2021reasonable}.
} of $J_n^{\text{opt}}$ for the study of gender discrimination by \citet{arceo-gomez2014race} (AGCV), where $J_n^{\text{opt}} \approx 2.65$. Under the null hypothesis that the true probabilities are consistent with the bivariate binomial mixture model, the bootstrap distribution suggests that much larger values than $J_n^{\text{opt}}$ can be realized under the null---its 95\% quantile is 9.9.

Building on their bootstrap analysis, we propose optimization problem (iv) to illustrate how uncertainty affects their bounds by allowing all distributions with $J_n(f_{\widetilde{G}}, \bar{f}) \leq \kappa$ for some choice of $\kappa >0$. For $\kappa < J_n^{\text{opt}}$, the problem is infeasible. When $\kappa = J_n^{\text{opt}}$ and assuming uniqueness of the minimizer, problems (iii) and (iv) yield identical optimal values. For $\kappa > J_n^{\text{opt}}$, problem (iv)'s optimal value can be strictly smaller, reflecting the incorporated additional uncertainty.

\begin{figure}
    \centering
    \begin{tabular}{cc}
        a) $\;\;\mathbb P_G[p_a > p_b \mid C_a=1, C_b=0]$ & b) $\;\;\mathbb P_G[p_a > p_b \mid C_a=4, C_b=0]$ \\ 
        \includegraphics[width=0.45\textwidth]{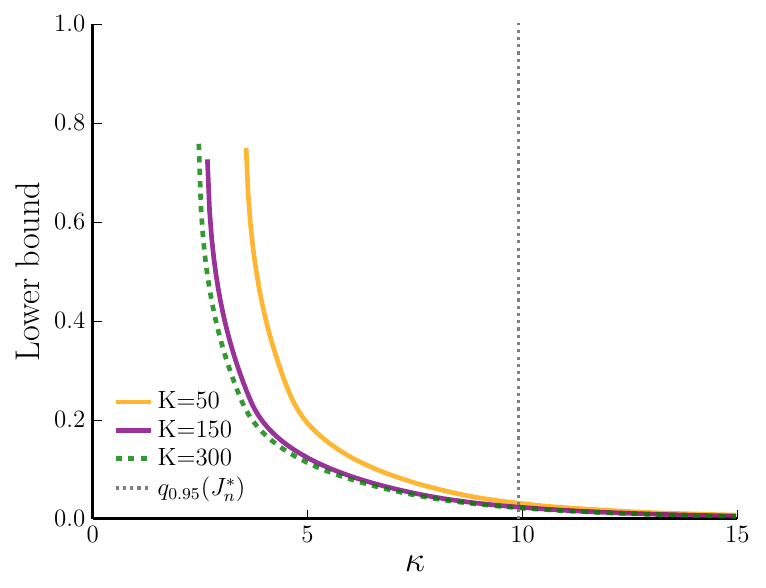} &
        \includegraphics[width=0.45\textwidth]{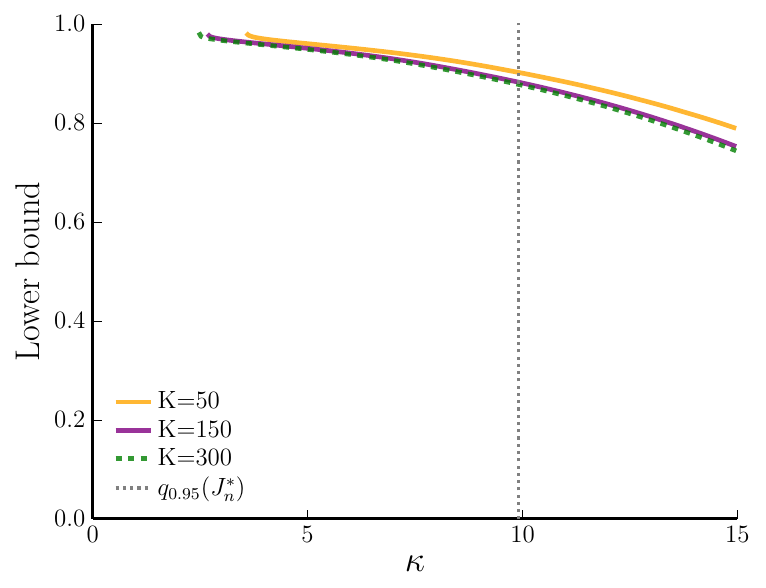}
    \end{tabular}
    \caption{Lower bounds as a function of the slack $\kappa$.}
    \label{fig:agcv_lower_bounds}
\end{figure}

Figure~\ref{fig:agcv_lower_bounds} shows the results of solving optimization problem (iv) for two callback patterns in the AGCV dataset and for different values of $\kappa$. In panel a) for $\theta^{\text{discr}}(G, (1,0))$ and in panel b) for $\theta^{\text{discr}}(G, (4,0))$.  For each pattern, we present results using three different discretization strategies corresponding to $K=50$, $150$, or $300$. All lower bound curves begin at $\kappa = J_n^{\text{opt}}$, with finer discretizations yielding slightly smaller values of $J_n^{\text{opt}}$. While discretization choices substantially impact the lower bounds when $\kappa$ is close to $J_n^{\text{opt}}$, these effects become less pronounced as $\kappa$ increases.

The analysis reveals stark differences in robustness across discrimination estimands for different callback patterns. \citeauthor{kline2021reasonable}' estimate that ``an employer that calls back a single woman and no men has at least a 74\% chance of discriminating against men'' proves highly sensitive to uncertainty---the lower bound rapidly drops as soon as we relax $\kappa$ beyond $J_n^{\text{opt}}$, falling to just 2\% at $\kappa = 9.9$ (the 95\% quantile of the bootstrap distribution of $J_n^{\text{opt}}$). By contrast, their estimate that ``at least 97\% of the jobs that call back four women and no men are estimated to discriminate against men'' demonstrates greater robustness, maintaining a lower bound of 88\% even at $\kappa = 9.9$. As explained in Section \ref{sec:$F$-Localization} below, this particular choice of $\kappa$ is a special case of the $F$-localization approach of \citet{ignatiadis2022confidence} and ensures that the lower bound is a valid 95\% lower confidence bound on $\theta(G;z)$. Table~\ref{tab:bounds_discr} records this lower confidence bound as well as lower confidence bounds from two other methods that are described in detail in Section \ref{sec:further_methods}.  Collectively, these results demonstrate the importance of accounting for sampling uncertainty in empirical Bayes discrimination analysis, as some findings prove substantially more robust than others.

\begin{table}[htbp]
    \centering
    \begin{tabular}{l|cc}
        \hline
        & $(C_a, C_b) = (1,0)$ & $(C_a, C_b) = (4,0)$ \\
        \hline
        $F$-Localization ($\kappa=9.9$) & 0.02 & 0.88 \\
        AMARI & 0.01 & 0.92 \\
        FSST & 0.01 & 0.89\\
        \hline
    \end{tabular}
    \caption{Lower bounds of 95\% confidence intervals for $\theta^{\text{discr}}(G;z)$ for $z=(1,0)$ and $z=(4,0)$.} 
    \label{tab:bounds_discr}
\end{table}

\section{On the principle of $F$-localization}
\label{sec:$F$-Localization}
As mentioned previously, setting $\kappa$ equal to the 95\% quantile of the bootstrap distribution of $J_n^{\mathrm{opt}}$ is a special case of the $F$-localization approach of \cite{ignatiadis2022confidence} and ensures that the lower bound obtained in this way is a valid lower confidence bound on $\theta(G;z)$.  To see why, 
suppose that we choose a potentially data-driven $\widehat{\kappa}$ such that $\mathbb P_G[J_n(f_{G}, \bar{f}) \leq \widehat{\kappa}] \geq 1-\alpha$. Fix an estimand of interest $\theta(G;z)$ and solve optimization problem~\eqref{eq:opt}(iv) at $\kappa = \widehat{\kappa}$ calling the optimal value $\underline{\theta}(\widehat{\kappa})$. It follows that
$$
\mathbb P_G[ \theta(G;z) \geq \underline{\theta}(\widehat{\kappa})] \geq  \mathbb P_G[J_n(f_{G}, \bar{f}) \leq \widehat{\kappa}] \geq 1-\alpha,
$$
where the first inequality follows since on the event $\{J_n(f_{G}, \bar{f}) \leq \widehat{\kappa}\}$, $G$ is a feasible solution in optimization problem~\eqref{eq:opt}(iv).

The above construction is a special case of the $F$-localization principle \citep{ignatiadis2022confidence}. The key idea is to construct a $(1-\alpha)$-confidence set of marginal distributions $\mathcal{F}(\alpha)$ such that $\mathbb P_G[F_G \in \mathcal{F}(\alpha)] \geq 1-\alpha$, where $F_G$ denotes the marginal distribution of $Z$ under $G$, i.e., the distribution with density $f_G$ defined in~\eqref{eq:marginal_density}. From this confidence set, one can construct confidence intervals for any functional of interest $\theta(G;z)$ by optimizing over all distributions $G$ whose marginals lie in $\mathcal{F}(\alpha)$, similar to the optimization problem in~\eqref{eq:opt}(iv).\footnote{
It is possible that there is no  $\widetilde{G} \in \mathcal{G}$ such that $F_{\widetilde{G}} \in \mathcal{F}(\alpha)$. 
If $\mathcal{F}(\alpha)$ is a valid $F$-localization, then this can happen for two reasons: we are on the low probability event that $F_G \notin \mathcal{F}(\alpha)$ or the model is misspecified. Thus the $F$-Localization approach includes an embedded specification test, similar to e.g., \citet{romano2008inference} and \citet{stoye2009more}.}

In this way, the $F$-localization approach translates statements concerning the uncertainty about the distribution of observables $F_G$ into statements concerning the uncertainty about the latent distribution $G$ and functionals thereof. The projection idea underlying the $F$-Localization approach traces back to the fundamental ideas of \citet{scheffe1953method} and \citet{anderson1969confidence}.
There are several ways of constructing $F$-localizations. For instance, one generic approach that works for any univariate $Z_i$ is to use the Dvoretzky-Kiefer-Wolfowitz inequality with Massart's~\citeyearpar{massart1990tight} tight constant. In other cases, more specialized and refined constructions can work instead; for instance, \citet{ignatiadis2022confidence} construct a $F$-localization in the Gaussian empirical Bayes problem by considering an $L_{\infty}$ neighborhood of the marginal density $f_G$. For discrete problems, such as the one considered by~\citeauthor{kline2021reasonable}, methods using a $\chi^2$-based $F$-localization were already developed by \citet{lord1975empirical,lord1976interval}.

An important feature of the $F$-localization principle is that it can be used to construct confidence intervals for any functional of the distribution $G$, provided that the resulting optimization problem is tractable. Moreover, all these confidence intervals have simultaneous $1-\alpha$ coverage: if $F_G \in \mathcal{F}(\alpha)$, which has probability at least $1-\alpha$ if $\mathcal{F}(\alpha)$ is a valid $F$-localization, then all the resulting confidence intervals will cover the true value of the functional with at least the desired probability. Simultaneity is a desirable property when the empirical Bayes analysis is highly exploratory as in~\citeauthor{kline2021reasonable}: therein the authors consider all kinds of estimands: $\theta^{\text{discr}}(G;z)$ for different callback patterns $z$; alternative definitions of discrimination, e.g., $\mathbb P_G[p_a \neq p_b \mid Z=z]$ (again, for different $z$); unconditional discrimination probabilities such as $\mathbb P_G[p_a > p_b]$ and so forth. The $F$-localization principle allows one to construct confidence intervals for all of these estimands with simultaneous coverage.

\subsection{Computational issues for $F$-localization}
\label{subsec:optimization_issues}

It is common in empirical Bayes problems to discretize the space of distributions $\mathcal{G}$ \citep{koenker2014convex}. For the discrimination detection problem, where $\mathcal{G}$ consists of distributions over $[0,1]^2$, we introduce a grid \smash{$\mathcal{D}_K = \{\theta_{\ell}\,:\,\ell=1,\dotsc,K^2\} \subset [0,1]^2$} (following~\citealp[Appendix C]{kline2021reasonable}) and represent distributions as \smash{$\widetilde{G} = \sum_{\ell=1}^{K^2} \pi_{\ell} \delta_{\theta_{\ell}}$}
where $\delta_{\theta_\ell}$ denotes the Dirac measure at $\theta_\ell \in [0,1]^2$, and the weights $\pi_{\ell}$ satisfy $\pi_{\ell} \geq 0$ and $\sum_{\ell} \pi_{\ell} = 1$. Ideally, one should use as fine a grid as computationally feasible. In Figure \ref{fig:agcv_lower_bounds}, we use, like \citeauthor{kline2021reasonable}, the above discretization scheme with varying grid sizes corresponding to $K=50$, $150$, $300$ to assess sensitivity to discretization.

Following \citeauthor{kline2021reasonable}, let us explain how this discretization turns optimization problems (i)--(iii) into linear programs. Common empirical Bayes estimands, including the discrimination estimand $\theta^{\text{discr}}(G;z)$, take the form,
\begin{equation}
\theta^{\text{post}}(G;z) = \mathbb{E}[h(\theta) \mid Z=z] = \frac{ \int h(\theta) p(z \mid \theta) \dd G(\theta)}{  f_G(z)},
\label{eq:theta_post}
\end{equation}
for a function $h(\cdot)$, e.g., $h(\theta) = \ind(p_a > p_b)$.
For such estimands, after discretization, one can solve optimization problem \eqref{eq:opt}(iii) (and analogously, (i) and (ii)) by linear programming: 
$$
\begin{aligned}
& \underset{\pi \in [0,1]^{K^2}}{\text{minimize}}
& &\sum_{\ell=1}^{K^2} h(\theta_{\ell})\frac{p(z \mid \theta_{\ell})}{{\bar{f}^{\text{proj}}(z)}}\pi_{\ell}
& \text{s.t.}
& & \sum_{\ell=1}^{K^2} p(z' \mid \theta_{\ell})\pi_{\ell} = \bar{f}^{\text{proj}}(z')\, \; \text{for all } z' \in \mathcal{Z},\quad\sum_{\ell=1}^{K^2} \pi_{\ell} = 1.
\end{aligned}
$$
Observe that the objective and the constraints are linear in the $\pi_{\ell}$.\footnote{
Note that $\bar{f}^{\text{proj}}$ is computed in a first step in a separate optimization problem and is treated as fixed in the linear program; by doing so, the ratio objective becomes linear in the optimization variables.  The fractional programming techniques we describe below allow directly solving optimization problems with a ratio objective.
} Analogously, after discretization, optimization problem~\eqref{eq:J_functional} is also a convex program that can be solved by second order conic programming (SOCP).\footnote{
Note that the first stage GMM matrix $\widehat{W}$ in~\eqref{eq:J_functional} and the bootstrap distribution of $J_n^{\mathrm{opt}}$ shown in Figure~\ref{fig:bootstrap}, also depend on the discretization. We ignore this dependence for simplicity and compute these quantities only under the $K=150$ grid.}

It turns out that we can substantially extend both the class of estimands and the constraints (beyond linear) and still maintain computational tractability that facilitates the construction of $F$-localization based confidence intervals. Concretely, consider any estimand that may be written as a ratio of linear functionals of $G$,
\begin{equation}
\theta^{\text{ratio}}(G;z) = \frac{ N(G;z)}{D(G;z)},
\label{eq:ratio_estimand}
\end{equation}
with $N(G;z)$ and $D(G;z)$ linear functionals of $G$.\footnote{For instance, the estimand in~\eqref{eq:theta_post} can be written in this way by setting $N(G;z) = \int h(\theta) p(z \mid \theta) \dd G(\theta)$ and $D(G;z) = f_G(z)$.}
Then, the optimization problem~\eqref{eq:opt}(iv) can also be solved as a SOCP using techniques from fractional programming~\citep{charnes1962programming}; see~\citet{ignatiadis2022confidence} for details. Hence, e.g., computing the lower bounds in Figure~\ref{fig:agcv_lower_bounds} for the discrimination estimand $\theta^{\text{discr}}(G;z)$ is computationally fast even for the grid with $300^2$ points.

\section{Inference methods beyond $F$-localization}
\label{sec:further_methods}
In some situations, because it achieves simultaneous coverage over all possible empirical Bayes estimands, $F$-localization may be overly conservative. Some recent innovations permit construction of confidence intervals that have nominal coverage for a specific estimand of interest, and so, in some cases, can be substantially shorter than $F$-localization intervals. Here, we describe two approaches, both of which account for both sources of uncertainty, partial identification and sampling variability. Discretization considerations for these methods are similar to those for $F$-localization.

\subsection{Affine Minimax Anderson-Rubin Inference (AMARI)} This method developed in~\citet{ignatiadis2022confidence} provides confidence intervals for any ratio estimand of the form in~\eqref{eq:ratio_estimand}. The starting point is to test for each $c$ whether $\theta^{\text{ratio}}(G;z)=c$ and to obtain a confidence interval by inversion. By an Anderson-Rubin-type argument, it thus suffices to test whether $L(G;c):=  N(G;z) - c D(G;z) = 0$, where $L$ is a linear functional of $G$. Given this reduction, the method proceeds by bias-aware inference using the affine minimax approach of \citet{donoho1994statistical} and \citet{armstrong2018optimal} carefully tailored to the empirical Bayes setting. AMARI requires a pilot $F$-Localization, and here we use the $F$-Localization implied by the constraint $J_n(f_{\widetilde{G}}, \bar{f}) \leq 13.2$ (where $\kappa=13.2$ is the $99\%$ quantile of the bootstrap distribution of $J_n^{\mathrm{opt}}$). We refer to~\citet{ignatiadis2022confidence} for more details.

\subsection{\cite{fang2023inference} (FSST)} This method was not developed for the empirical Bayes setting per say, yet here we observe that it is applicable to discrete empirical Bayes problems, such as the one here with a bivariate binomial likelihood. Using the same Anderson-Rubin-type argument as above, the confidence interval for $\theta^{\text{ratio}}(G; z)$ is the collection of values of $c$ for which the null hypothesis that $L(G;c) = 0$ can be not rejected. Since $L(G;c)$ is a linear functional of $G$, after discretization as described above, this null hypothesis can be restated as 
\[
\exists \pi \in \mathbb R^d_+ \text{ such that } A\pi = \beta \text{ and } a'\pi = 0~,
\]
where $d = K^2$, $\beta = f_G$, $A$ is a $d \times p$-dimensional matrix that encodes the bivariate binomial likelihood function, evaluated at different $z \in \mathcal{Z}$ and the grid of $K^2$ elements $(p_a,p_b) \in [0,1]^2$, and $a'\pi$ encodes the restriction that $L(G;c) = 0$. \citet{fang2023inference} develop a general approach to testing such a null hypothesis.  See also \cite{bai2024generalizedstrata} for related applications of this methodology in causal inference.

\subsection{Further methods} There are further potential ways in which one can form confidence intervals, for instance, by pursuing the Anderson-Rubin-type argument above and test inversion. For instance, we could use CNS again (which above we used to facilitate $F$-Localization) to test the null hypothesis $L(G;c)=0$, see~\citet[Remark 2.3]{chernozhukov2023constrained}. Yet another alternative (that however, in general, lacks distribution-uniform coverage) is given in~\citet{d2017measuring}.

\section{On the choice of estimand: a counterfactual odds ratio} \label{sec:estimand}

Building on our previous analysis of uncertainty quantification, we now turn to a fundamental question that underlies the entire empirical Bayes approach to discrimination detection: the choice of estimand itself. The discrimination estimand in~\eqref{eq:discrimination_estimand} presents two challenges that intertwine with our previous discussion of uncertainty.  First, the estimand $\theta^{\text{discr}}(G;z)$ is discontinuous with respect to weak convergence of measures. This discontinuity complicates interpretation as small perturbations in the ensemble $G$ can lead to large changes in this discrimination estimand.\footnote{
A similar critique also applies to common multiple testing analyses. For instance, the critique applies to the local false discovery rate $\theta^{\text{lfdr}}(G;z) := \mathbb P_G[\theta = 0 \mid Z=z]$ in the Gaussian empirical Bayes problem with $\theta \sim G$, $Z \mid \theta \sim \mathrm{N}(\theta, 1)$~\citep{mccullagh2018statistical,xiang2024interpretation}. 
} Second, it does not reflect the magnitude of discrimination, which, in practical policy applications, is often relevant for resource allocation and enforcement prioritization.  To illustrate these concerns, consider a distribution $G$ where $p_a = p_b + 10^{-10}$ almost surely. Then $\theta^{\text{discr}}(G;z) = 1$ for all $z$, suggesting complete discrimination against group $b$, even though such a small difference would not lead to any observable differences in hiring patterns. Moreover, if we slightly perturb $G$ such that $p_a = p_b$ almost surely, then $\theta^{\text{discr}}(G;z) = 0$ for all $z$. We note that while the exact-zero discrimination threshold creates technical challenges for the estimand, this binary framing aligns with certain legal frameworks such as the Civil Rights Act, where discrimination of any magnitude is in violation of the law.

Motivated by such concerns, \citet[Lemma 3]{kline2021reasonable} propose the logit estimand,\footnote{\citet{kline2021reasonable} also incorporate applicant quality in their estimand definition, which we omit.} 
$$\theta^{\text{logit}}(G; z) := \mathbb E_G\left[\Lambda\left(\Lambda^{-1}(p_a) - \Lambda^{-1}(p_b) \right) \mid Z=z\right],\quad \Lambda(p):=\frac{\exp(p)}{1+\exp(p)}.$$
The logit estimand captures differences between group callback probabilities. Building upon this foundation, we propose a complementary counterfactual estimand that offers an alternative perspective on measuring discrimination magnitude.  Our estimand  answers the following question: ``If we were to send additional applications to an employer with callback pattern $z=(c_a,c_b)$, what is the relative probability of observing strictly more callbacks for group $a$ versus group $b$?'' Formally, for employer $i$ with observed callback pattern $(C_a,C_b)=(c_a,c_b)$, consider the counterfactual experiment of sending $L'$ additional applications from each group, resulting in callbacks $C_a'$ and $C_b'$. Note that $L'$ could be different from $L$ sent in the original experiment. To define counterfactual probabilities, we assume that conditional on $\theta = (p_a, p_b)$, $C_a'$ and $C_b'$ are independent of $C_a$, $C_b$, and follow the bivariate binomial in~\eqref{eq:bivariate_binomial} with $L'$ trials. In this sense, we assume that our new experiment is a perfect replication of the original one except for a potentially different number of applications.  See \citet{yang2024largescale} for a related notion of an idealized replication experiment. With this setup, we define the ``posterior callback odds ratio'' given $z=(c_a,c_b)$ as
\begin{equation}
    \theta^{\text{odds}}(G; z, L') := \frac{ \mathbb P_G[ C_{a}^{'} > C_b'\mid C_a=c_a,C_b=c_b]}{\mathbb P_G[ C_{a}^{'} < C_b'\mid C_a=c_a,C_b=c_b] }.
\label{eq:new_estimand}
\end{equation}
This estimand represents the odds ratio of callbacks for group $a$ versus group $b$ in a counterfactual experiment that the experimenter could actually implement. It has a natural betting interpretation: it quantifies the odds one would accept when wagering that group $a$ will receive strictly more callbacks than group $b$ (rather than strictly fewer) in a new experiment, given the observed callback pattern. For instance, if $\theta^{\text{odds}}(G; z, L') = 3$, a rational decision-maker would be willing to bet up to 3:1 odds on group $a$ receiving more callbacks than group $b$ in a counterfactual experiment with $L'$ applications per group. This provides a meaningful quantification of discrimination that is tied to  outcomes.

This estimand in~\eqref{eq:new_estimand} has several desirable properties, as we next document (see Appendix~\ref{sec:prop_proof} for a proof).

\begin{proposition}[Properties of posterior callback odds ratio]
\label{prop:odds}
Let $\theta^{\text{odds}}(G; z, L')$ be defined as in~\eqref{eq:new_estimand}. Then:
\begin{enumerate}[label=(\alph*)]
    \item (No-discrimination baseline.) If $p_a=p_b$ almost surely under $G$, then $\theta^{\text{odds}}(G; z, L')=1$ for all callback patterns $z=(c_a,c_b)$.
    \item (Continuity under weak convergence.) If $G_n \cd G$ weakly, then $\theta(G_n;z,L') \to \theta(G;z,L')$.
    \item (Asymptotics with increasing number of applications $L'$.) As $L' \to \infty$:
    $$\theta(G;z,L') \to \frac{\mathbb P_G[ p_{a} = p_b \mid Z=z]/2 \,+\, \mathbb P_G[ p_{a} > p_b \mid Z=z]}{\mathbb P_G[ p_{a} = p_b \mid Z=z]/2 \,+\,\mathbb P_G[ p_a < p_b \mid Z=z]},$$
    with the convention that the right hand side is $\infty$ when its denominator is zero. 
    In words, if we could send infinitely many applications in our counterfactual experiment, then the odds ratio estimand in~\eqref{eq:new_estimand} can be interpreted as a dampened ratio of discrimination probability $\theta^{\text{discr}}(G;z)$ in~\eqref{eq:discrimination_estimand} for group $a$ versus group $b$ divided by the discrimination probability for group $b$ versus group $a$.
\end{enumerate}
\end{proposition}
An important property of $\theta^{\text{odds}}(G; z, L')$ is that it can be expressed as a ratio of linear functionals of $G$:
\begin{equation}
\label{eq:odds}
\theta^{\text{odds}}(G;(c_a,c_b),L') = \frac{\int_{[0,1]^2}\sum_{c_b'=0}^{L'-1} 
 \sum_{c_a'=c_b'+1}^{L'} p(c_a',c_b'\mid p_a,p_b) p(c_a, c_b \mid p_a, p_b) \dd G(p_a,p_b)}{ \int_{[0,1]^2}\sum_{c_b'=1}^{L'} 
 \sum_{c_a'=0}^{c_b'-1} p(c_a',c_b'\mid p_a,p_b) p(c_a, c_b \mid p_a, p_b) \dd G(p_a,p_b) }.
\end{equation}
Hence, this estimand is amenable to uncertainty quantification methods including $F$-localization described in Section~\ref{sec:$F$-Localization} and other methods described in Section~\ref{sec:further_methods}. Table~\ref{tab:bounds} presents confidence intervals for this estimand using $L'=4$ on the AGCV dataset. For the pattern $(C_a,C_b)=(1,0)$, all three inference methods yield intervals containing 1, suggesting insufficient evidence of systematic discrimination. In contrast, for $(C_a,C_b)=(4,0)$, all methods yield intervals strictly above 1, with AMARI providing a tighter lower bound of 17. The interpretation is operationally clear: if we send 4 more applications, we are much more likely to observe a callback pattern favoring the first applicant group.

\begin{table}[htbp]
    \centering
    \begin{tabular}{l|cc}
        \hline
        & $(C_a, C_b) = (1,0)$ & $(C_a, C_b) = (4,0)$ \\
        \hline
        $F$-Localization ($\kappa=9.9$) & 0.62 & 8.5 \\
        AMARI & 0.51 & 17.0 \\
        FSST & 0.41 & 8.9 \\
        \hline
    \end{tabular}
    \caption{Lower bounds of 95\% confidence intervals for different methods for the posterior callback odds ratio estimand in~\eqref{eq:new_estimand} with $L'=4$ and initial callback patterns $(C_a,C_b)=(1,0)$, respectively, $(C_a, C_b) = (4,0)$.
    }
    \label{tab:bounds}
\end{table}

Lastly, we note that the posterior estimand is often used to inform judgments and policy decisions regarding firms with a specific callback pattern (e.g., deciding which firms to audit or sanction). For further discussion, see the work on firm discrimination in \cite{kline2024discrimination}, as well as related practices in teacher value-added models \citep{gilraine2020new} and medical facility rankings \citep{gu2023invidious}. The common empirical Bayes estimand typically takes the form of a posterior expectation, which ensures, for a suitable choice of loss function, that decisions are of high-quality on average. In the specific setting of \citet{kline2021reasonable}, suppose $\theta^{\text{discr}}(G; z) = 0.99$, then if a policy maker were to audit all employers (or a random subset thereof) having this callback pattern $z$, only 1\% of the resources would be spent on non-discriminating firms. However, for decisions of particularly high stakes, e.g., imposing sanctions on individual firms, the above criterion may not be sufficiently conservative and one may prefer a frequentist approach that provides error control for individual firms without relying on the exchangeability of all firms. See, e.g., the methods developed in \citet{mogstad2024inference} and the related discussion in \citet{mogstad2022comment}.  Even so, the empirical Bayes approach may provide useful preliminary evidence.  In such use, as emphasized in our discussion above, it is additionally important to account for uncertainty from partial identification and sampling.

\appendix 

\section{Proof of Proposition~\ref{prop:odds}}
\label{sec:prop_proof}
\begin{proof}

Part (a) follows by iterated expectation and noting that $C_a'$ is iid with $C_b'$ conditional on any value of $\theta$ on the support of $G$:
$$\mathbb P_G[ C_{a}^{'} > C_b'\mid Z=z] = \mathbb E_G[ \mathbb P[ C_{a}^{'} > C_b'\mid \theta] \mid Z=z] =\mathbb E_G[ \mathbb P[ C_{a}^{'} < C_b'\mid \theta] \mid Z=z]   = \mathbb P[ C_{a}^{'} < C_b'\mid Z=z].$$

For part (b) we may argue via representation~\eqref{eq:odds} and proving convergence for the numerator and denominator separately. Call the numerator $N(G)$, omitting explicit dependence on $z, L'$ and observe that $N(G) = \int \psi(p_a, p_b) \dd G(p_a, p_b)$, where $\psi$ is a polynomial and thus bounded and continuous on $[0,1]^2$. It follows that $N(G_n) \to N(G)$ when $G_n \cd G$. The argument for the denominator is analogous.

For part (c), we write $\theta^{\text{odds}}(G; z, L') = \mathbb P_G[ C_{a}^{'} > C_b', Z=z]/\mathbb P_G[ C_{a}^{'} < C_b',Z=z]$ and again argue about the limits of the numerator and denominator separately. For the numerator, it holds that:
$$
\mathbb P_G[ C_{a}^{'} > C_b', Z=z] = \int \mathbb P[ C_{a}^{'} > C_b', Z=z \mid \theta]  \dd G(\theta) = \int \mathbb P[ C_{a}^{'} - C_b' > 0 \mid \theta] \mathbb P[ Z=z \mid \theta]  \dd G(\theta)
$$
Now notice that by the central limit theorem,
$$ \lim_{L' \to \infty} \mathbb P[ C_{a}^{'} - C_b' > 0 \mid \theta] = \begin{cases} 1,\,\,\,\;\,\,\, \text{if}\,\,\, p_a  > p_b, \\ 
0,\,\,\,\;\,\,\, \text{if}\,\,\, p_a < p_b,\\ 
1/2,\,\, \text{if}\,\,\, p_a = p_b.\end{cases}
$$
By dominated convergence, it follows that 
$$\mathbb P_G[ C_{a}^{'} > C_b', Z=z] \to \int \{( \ind(p_a=p_b)/2 + \ind(p_a > p_b)\} \mathbb P[ Z=z \mid \theta]  \dd G(\theta),$$
and the right hand side the same as $\{\mathbb P_G[ p_a=p_b \mid Z=z]/2 + \mathbb P_G[p_a > p_b \mid Z=z]\} \mathbb P_G[Z=z]$. We argue analogously regarding the denominator, and may so conclude.
\end{proof}

\subsection*{Acknowledgments.} We thank Chris Walters for helpful feedback on an earlier version of this manuscript. 

\bibliography{Flocal}

\end{document}